\documentstyle[aps]{revtex}
\input{epsfig.sty}

\def\al{\alpha}
\def\beq{\begin{equation}}
\def\eeq{\end{equation}}
\def\bea{\begin{eqnarray}}
\def\eea{\end{eqnarray}}
\def\lsim{\mathrel{\rlap{\lower4pt\hbox{\hskip1pt$\sim$}}
    \raise1pt\hbox{$<$}}}         
\def\gsim{\mathrel{\rlap{\lower4pt\hbox{\hskip1pt$\sim$}}
    \raise1pt\hbox{$>$}}}
\begin{document}
\title{Bubble collisions in a SU(2) model of QCD}
\author{
Mikkel B. Johnson \\
Los Alamos National Laboratory, Los Alamos, NM 87545 \\
 Ho-Meoyng Choi \\
Department of Physics, Carnegie-Mellon University,
Pittsburgh, PA 15213\\
and Department of Physics, Kyungpook National University, Taegu,
702-701 Korea\\
Leonard S. Kisslinger\\
Department of Physics, Carnegie-Mellon University,
Pittsburgh, PA 15213 }
\maketitle
\begin{abstract}
Starting from the QCD action in an instanton-like SU(2) Yang-Mills
field theory, we derive equations of motion in Minkowski space. 
Possible bubble collisions are studied in a $(1+1)$-dimension reduction. 
We find gluonic structures which might give rise to CMBR effects.
\vspace{3mm}

PACS Indices:12.38.Lg,12.38.Mh,98.80.Cq,98.80Hw
\vspace{3mm}


\end{abstract}


\section{Introduction}

   In the early universe during the time interval between 
10$^{-5}$-10$^{-4}$sec the universe passed through the temperature (T) range
which included the critical T=T$_c$ for the QCD chiral phase transition from 
the quark-gluon plasma (QGP) to the hadronic phase (HP), T$_c\simeq$ 150 MeV. 
This is the quark-hadron phase transition (QHT). In the present work we assume 
that this phase transition is first order, in which case bubbles of the HP 
could form, nucleate and collide until the our HP universe emerges with T at 
about 100 MeV at 10$^{-4}$s. Although at the present time lattice calculations 
cannot determine the order of the transition\cite{lat1}, some lattice gauge 
calculations\cite{lat2} and numerical model calculations\cite{nm} indicate 
that the transition is weakly first order, which would imply bubble 
formation during the QHT.

   The crucial question is whether the phase transition leads to astrophysical 
observables. Based on the results of an effective field model in which
an internal QCD domain wall is produced within the HP bubble during the QHT 
and results in a magnetic wall\cite{fz1}, a study of possible Cosmic Microwave
Background Radiation (CMBR) correlations has been carried out\cite{lsk1}.
It was found that the large-scale magnetic wall could lead to CMBR polarization
correlations that might be observable in the near future. This result depends
on the existence of a domain-size thin gluonic wall with a lifetime long enough
to produce the magnetic wall. The latter continues to exist and evolves until
the last scattering time (about 300,000 years), and can therefore lead to CMBR
polarization and metric correlations.

   Although there has been a great deal of work on bubble nucleation during
the QHT\cite{ign}, a subject that is still incomplete, there has been very 
little published on the bubble collisions. 
For the electroweak phase transition (EWT)
bubble collisions were studied two decades ago\cite{hms} using a Higgs model 
and a semiclassical tunneling picture that was originally developed for 
nonperturbative QCD\cite{col}. In recent years an Abelian Higgs model\cite{kv}
has been used to study EWT bubble collisions\cite{kv,ae,cst} and possible
magnetic fields which are produced in such collisions.

   In the present paper bubble collisions are studied for the QHT using an
instanton-inspired model of QCD. Instantons can be used to represent the 
midrange nonperturbative QCD interactions, and recently have been used
to study high-energy collisions of quark/gluon systems\cite{nowak}. Also,
for the QHT an instanton model for the bubble wall between the QGP and
 HP is consistent with the surface tension found in lattice gauge 
calculations\cite{lsk2}. However, the bubble nucleation and collisions
have not been calculated in such a picture. We do not attempt the study
of nucleation in this work, but center on investigation of the possible
production of an interior gluonic wall, which could possibly lead to the
CMBR correlations found in Ref.\cite{lsk1}
    
  The equations of motion for the QCD pure-gluonic Lagrangian, i.e., the
equations of motion for the SU(3) color field without quarks, are given in
Sec. II, with a brief discussion of the instanton model. The equations of 
motion for the instanton-motivated SU(2) model of QCD used in the present 
work, which we call SU(2) Yang-Mills fields to distinguish the model from 
SU(3) QCD, are given in Sec III. The instanton-type form is in 
4-dimensional Euclidean space, which we continue to (3+1)-dimensional 
Minkowski space for the investigation of bubble collisions; and we choose
a convenient metric, which we discuss in detail. The 
energy-momentum tensor in Minkowski space is also derived. 
In Sec. IV we study (1+1)-dimensional collisions in 
Minkowski space and find promising gluonic structure arising in the interior.
This could be the basis for investigating large-scale structure that could
lead to CMBR correlations or large-scale galaxy effects. Our conclusions are
discussed in Sec. V.

\section{SU(3) color field equations of motion and instanton approximation}
   In this section we consider the purely gluonic part of the QCD Lagrangian,
derive the field equations of motion, and discuss the relation to the
instanton model.  Since in the present work we do not solve these equations, 
but use the SU(2) instanton-motivated version described in the next section,
we only give a brief discussion here.

   The Lagrangian density for pure glue is
\bea
\label{glue}
  {\cal L}^{glue} & = & \frac{1}{4} G \cdot G
\eea
with
\bea
\label{G}
    G_{\mu\nu} & = & \partial_\mu A_\nu -  \partial_\nu A_\mu
-i g [A_\mu,A_\nu]\\
    A_\mu & = & A_\mu^n \lambda^n/2 \nonumber
\eea
with $\lambda^n$ the eight SU(3) Gell-Mann matrices, $([\lambda_a,\lambda_b]
=2if_{abc}\lambda_c)$. Minimizing the action,
\bea
\label{action}
 S & = &\frac{1}{4}\int\; d^4x \sum_a G^a_{\mu\nu}G^{\mu\nu a},
\eea
one obtains equations of motion (EOM) for the color field
\bea
\label{qcdeom1}
    \partial_\mu G^{\mu\nu a} +g f^{abc} A^b_\mu G^{\mu\nu c} & = & 0.
\eea.

Using the Lorentz gauge, with the gauge condition $\partial_\mu A^a_\mu=0$ 
in Eq. (\ref{qcdeom1}) one obtains the EOM for the SU(3)
color field
\bea\label{qcdeom2}
\partial_\mu\partial^\mu A^a_\nu 
+ g f^{abc}(2 A^b_\mu\partial^\mu A^c_\nu 
- A^b_\mu \partial_\nu A^{\mu c}) 
+ g^2 f^{abc} f^{cef}A^b_\mu A^{\mu e}A^f_\nu
= 0,
\eea 
where the roman($a,b,c,\cdots$) and greek($\mu,\nu$) indices
run from 1 to 8 and from 1 to 4, respectively. An approximate solution
to the equations of motion is found by using the instanton approximation
for the QCD Lagrangian density\cite{bev}, in which a classical SU(2)
Yang-Mills field is used for the color field and the minimization of the
action,  $\delta S= \delta \int d^4x {\cal L} ^{instanton} = 0$, for a
pure gauge field in Euclidean space has the solution
\bea
\label{inst}
 A_\mu^{n,inst}(x)& = & \frac{2 \eta^{-n}_{\mu\nu}x^\nu}{(x^2 + \rho^2)}\\
        G^{n,inst}_{\mu\nu}(x) & = & -\frac{\eta^{-n}_{\mu\nu} 4 \rho^2}
{(x^2 + \rho^2)^2},
\eea
for the instanton and a similar expression with -n for the anti-instanton, 
where $\rho$ is the instanton size and the  $\eta^{n}_{\mu\nu}$ 
are defined in Ref.\cite{hooft}. The instanton connects points in two QCD 
vacua which differ by one unit of winding number. For our system one point
is in the QGP and the other in the HP. The model is discussed in Ref.\cite{ss},
and the extension to finite T has also been studied\cite{chu}.

   In the model used in the present work we start with the SU(2) gauge 
field as in the instanton model, and model the color field with the 
color/Dirac structure of the instanton solutions\cite{hooft}, but keep a more 
general space-time structure. The resulting equations of motion
starting with the general form of Eq.(\ref{qcdeom2}) are given in Sec III.
The instanton-type form is in 4-dimensional Euclidean space, which we continue
to (3+1)-dimensional Minkowski space. We find it convenient to use a
metric $g_{\mu\nu} = \delta_{\mu\nu}(1,1,1,1)$ in Euclidean space, so that
our Metric in Minkowski space is not the standard one. Of course this does
not alter our results, as we discuss in the next section.

\section{SU(2) Yang-Mills field equations of motion in Minkowski Space }
   Since we are using an instanton-inspired picture, we formulate the theory
in Euclidean space with the metric tensor  $g_{\mu\nu}={\rm diag}(1,1,1,1)$.
The analytic continuation to Minkowski space is made by $\tau \rightarrow
it$, so that the Minkowski space metric is  $g_{\mu\nu}={\rm diag}(-1,1,1,1)$,
with an overall sign change from the usual metric. We discuss any resulting
changes in our equations with this choice of metric. The SU(2) color gauge 
field is 
\bea
\label{ymfield}
A_{\mu} & = & A^a_{\mu}(\sigma^a/2),
\eea
where $\sigma_i$ is the Pauli
matrix satisfying $[\sigma^a,\sigma^b]=2i\epsilon^{abc}\sigma^c$
and Tr($\sigma^a\sigma^b$)$=2\delta^{ab}$. The field tensor is
\bea
\label{ymtensor}
    F_{\mu\nu} & = & F^a_{\mu\nu}(\sigma^a/2) \\ \nonumber
    F^a_{\mu\nu}& = & \partial_\mu A^a_\nu - \partial_\nu A^a_\mu
+ g\epsilon_{abc}A^b_\mu A^c_\nu,    
\eea
and the action, $S^{SU(2)}$, is
\bea
\label{eq1}
 S^{SU(2)} & = &\frac{1}{4}\int\; d^4x \sum_a F^a_{\mu\nu}F^{\mu\nu a},
\eea
This differs from the SU(3) $G_{\mu\nu}$ field tensor by the formal
replacement $f^{abc} \rightarrow \epsilon^{abc}$. Thus in the Lorentz
gauge, with the gauge condition $\partial_\mu A^a_\mu=0$ the equations of
motion for the SU(2) field follow from Eq.(\ref{qcdeom2}) with
$f^{abc} \rightarrow \epsilon^{abc}$, giving
\bea\label{EOM_A}
\partial_\mu\partial^\mu A^a_\nu 
+ g\epsilon^{abc}(2 A^b_\mu\partial^\mu A^c_\nu 
- A^b_\mu \partial_\nu A^{\mu c}) 
+ g^2\epsilon^{abc}\epsilon^{cef}A^b_\mu A^{\mu e}A^f_\nu = 0.
\eea 

   In the instanton model one divides the color field into the classical 
instanton field, and quantum fluctuation such as $A_\mu = A^{inst}_\mu + 
A^{qu}_\mu$. Since we shall solve the equations of motion for the bubble 
walls and wall collisions we cannot use the instanton solution directly, 
but we use an instanton-like form for the color field. In Euclidean space 
we take as our form for the color field (see Eq.(\ref{inst}))
\bea\label{A_inst}
A^{a }_\mu(x) &=& \frac{2}{g}\eta_{a\mu\nu}x^\nu F(x^2)
\equiv \eta_{a\mu\nu} W^\nu,
\eea
where the $\eta_{a\mu\nu}$ are defined in Ref.\cite{hooft}.

The Lagrangian density in terms of $W^\mu$ and $F(x^2)$ is given
by \footnote{
For a self consistency check, Eq. (\ref{LDW}) should give
$(F^a_{\mu\nu})^2=\frac{192\rho^4}{g^2(x^2+\rho^2)^4}$ for a
particular instanton solution $F=1/(x^2+\rho^2)$\cite{ss}.} 
\bea\label{LDW}
{\cal L}^{glue}&=&\frac{1}{4}F^a_{\mu\nu}F^{\mu\nu a}\nonumber\\
&=&\frac{1}{2}\biggl[
2(\partial_\mu W_\nu)^2 + (\partial_\mu W_\mu)^2
+ 4g(W^\mu W^\nu\partial_\mu W_\nu - W^2\partial_\mu W^\mu)
+ 3g^2 W^4 
\biggr]
\nonumber\\
&=&\frac{6}{g^2}\biggl[ x^2(\partial_\mu F)^2 + 4 F x^\mu\partial_\mu F
+ 8 F^2 - 8 x^2 F^3 + 4 x^4 F^4
\biggr],
\eea
where we use the following relations
\bea\label{ID2}
x^\mu\partial_\mu F(x^2) &=& 2 x^2\frac{\partial F(x^2)}{\partial x^2},
\nonumber\\
x^2\partial_\mu F &=& x_\mu x^\rho\partial_\rho F,
\eea
which follows from the fact that F depends only on $x^2$,
in deriving Eq. (\ref{LDW}).

Then the EOM in Eq. (\ref{EOM_A}) is given in terms of $W_\mu$ as
follows
\bea\label{EOM_W}
\partial^2 W_\mu &=& 2 g^2 W^2 W_\mu - 2 g W_\mu(\partial_\al W^\al)
+ 2 g W_\al \partial_\mu W^\al, 
\eea
with the gauge condition  $\partial_\mu A^a_\mu=0$ becoming
\bea
\label{gcondz}
   \epsilon_{\mu\nu\alpha\beta}\partial^\mu W^\nu & = & \partial_\beta W_\alpha
- \partial_\alpha W_\beta ,
\eea
and from Eq. (\ref{EOM_W}), we get
\bea\label{EOM_F2}
\partial^2 F + 12 F^2 - 8 x^2 F^3 + \frac{2}{x^2}x^\mu\partial_\mu F
&=& 0.
\eea 
Note that with our choice of metric (1,1,1,1) in Euclidean space,
$\partial^2 = \partial^2_{\vec x} + \partial^2_\tau 
( = \partial^2_{\vec x} - \partial^2_t)$ and  
$x^2={\vec x}^2 + \tau^2(={\vec x}^2-t^2)$ in 
Euclidean(Minkowski) space. From Eqs.(\ref{ID2},\ref{EOM_W},\ref{EOM_F2})
one can see that the standard choice (-1,-1,-1,-1), (1,-1,-1,-1)
for the Euclidean, Minkowski metrics would just change the sign of the
Function F, or result in the color fields getting a phase factor $e^{i\pi}$,
which has no physical consequence.

From Eq.(\ref{ID2})
\bea\label{ID_F}
{\vec x}^2 \partial_t F  &=& - t x^i\partial_i F,
({\rm or}\;\;
{\vec x}^2 \partial_\tau F  = \tau x^i\partial_i F).
\eea
Substituting Eq. (\ref{ID_F}) into Eq. (\ref{EOM_F2}), we obtain
\bea\label{EOM_F3}
\partial^2 F &=& -4\frac{\partial F}{\partial x^2}
 - 12 F^2  + 8 x^2 F^3.
\eea
or 
\bea\label{EOM_F4}
\frac{d^2 F}{d x^2}  + \frac{5}{x}\frac{d F}{dx}
&=& -12 F^2 + 8 x^2 F^3.
\eea
In the thin-wall approximation \cite{Lin1}, the first derivative term 
in Eq. (\ref{EOM_F4}) may be neglected.

The energy-momentum tensor is given by\footnote{
One could easily check that the energy density in Euclidean 
metric $T^{44}=0$ for $F=1/(x^2+\rho^2)$.}
\bea\label{EMT}
T^\mu_\nu&=& - \sum_a(F^{\mu\rho a} F^a_{\rho\nu}
-\frac{1}{4}g^\mu_\nu F^{\rho\sigma a} F^a_{\rho\sigma}),
\nonumber\\
&=& 2(\partial^\mu W_\nu)(\partial^\rho W_\rho)
+g^\mu_\nu(\partial^\rho W_\alpha)^2
- 2g\biggl[(\partial^\mu W_\nu)W^2 \nonumber\\
&&\;\;- (W^\mu W_\nu - g^\mu_\nu W^2)(\partial_\rho W^\rho)
 - g^\mu_\nu W^\rho W_\alpha\partial_\rho W^\alpha \biggr]
\nonumber\\
&&\;\; +2g^2(g^\mu_\nu W^4 - W^\mu W_\nu W^2) + g^\mu_\nu {\cal L}^{glue},
\nonumber\\  
&=& \frac{4}{g^2}\biggl[
2x^2(\partial^\mu F)(\partial_\nu F) + g^\mu_\nu x^2(\partial_\rho F)^2
+ 8F x^\mu\partial_\nu F + 4 g^\mu_\nu F x^\rho\partial_\rho F 
\nonumber\\
&&\;\;+ 12g^\mu_\nu F^2+ 16(x^\mu x_\nu - g^\mu_\nu x^2)F^3
+ 8(g^\mu_\nu x^2 - x^\mu x_\nu)x^2 F^4\biggr]
\nonumber\\
&&\;\;+ g^\mu_\nu{\cal L}^{glue}.
\eea

\section{Bubble Collisions in (1 + 1) dimension}
In this section we will study the EOM for the instanton solution
in (1+1) dimensional Minkowski space. This involves the replacement of
the Euclidean metric tensor 
$g^E_{\mu\nu}={\rm diag}(x_1,x_4)={\rm diag}(1,1)$ by the
Minkowski metric tensor $g_{\mu\nu}={\rm diag}(x_0,x_1)={\rm diag}(-1,1)$, 
and the analytic
continuation  $x_4\to ix_0$, mentioned above. Taking  $W_2=W_3=0$ and working
in the Minkowski space($x^0=t,x^1=x$),
the EOM given by Eq. (\ref{EOM_W}) is reduced to 

\bea\label{eom1}
 \partial^2 W_0(x,t) &=& 2 g^2 W_0(x,t)[W^2_1(x,t) - W^2_0(x,t)] - 
 2 g W_0(x,t)\partial_x W_1(x,t) \nonumber \\
&&\;\;+ 2 g W_1 \partial_t W_1(x,t),
\nonumber\\
  \partial^2 W_1(x,t) &=& 2 g^2 W_1(x,t)[W^2_1(x,t) - W^2_0(x,t)] + 
2 g W_1(x,t)\partial_t W_0(x,t) \nonumber \\
&&\;\; - 2 g W_0 \partial_x W_0(x,t),
\eea 
where $\partial^2 = \partial^2_x - \partial^2_t$ 
and $x_\mu x^\mu = x^2 - t^2$.
Note also that the EOM are constrained by the gauge condition.
The gauge condition Eq.(\ref{gcondz}) in 1+1 can be written
\bea
\label{gcondx}
             \partial_\mu W_\nu & = & \partial_\nu W_\mu.
\eea
In our 1+1 calculations the constraint on the t-derivative in Euclidean 
space is:
\bea
\label{conditionE}
      \partial_\tau F & = & \frac{\tau}{x^2} x \partial_x F,   
\eea
while in 1+1 Minkowski space, with $\tau^2=-t^2$ and $x^\mu x_\mu = x^2-t^2$,
with the notation $x^1 = x$ and $x^0 = t$, 
the gauge condition is
\bea
\label{conditionM}
       t \partial_t F & = & - \frac{t^2}{x^2} x \partial_x F.
\eea
Using the condition given by 
Eq.~(\ref{conditionM}) with the Minkowski form for Eq.~(\ref{EOM_F2}) one
obtains the two equivalent equations for $F(x,t)$ in 1+1 space
\bea
\label{eominst}
   \partial^2 F & = & -\frac{2}{x}\partial_x F -12 F^2 +8 (x^2-t^2)F^3\\
\nonumber
   \partial^2 F & = &  \frac{2}{t}\partial_t F -12 F^2 +8 (x^2-t^2)F^3.
\eea
Having solved for $F(x,t)$ one can obtain the energy density, $T_{00}$,
which was used to fit the bubble surface tension in Ref.~\cite{lsk2}, from 
Eqs.~(\ref{LDW},\ref{EMT}). The gluonic Lagrangian density is given by
\bea\label{Lglue}
{\cal L}^{glue} &=&\frac{1}{2}\biggl\{
 3 (\partial_t W_0)^2 + 3 (\partial_x W_1)^2 - 2(\partial_t W_1)^2
-2 (\partial_x W_0)^2
+ 4g[W^2_1\partial_t W_0 + W^2_0\partial_x W_1 
\nonumber\\
&&\;\; - W_0W_1(\partial_t W_1 + \partial_x W_0)]
+ 3 g^2 (W^2_1 - W^2_0)^2 
\biggr\}
\nonumber\\
&=& \frac{6}{g^2}\biggl[\frac{(x^2-t^2)^2}{x^2}(\partial_x F)^2
 +4\frac{(x^2-t^2)}{x}F \partial_x F+ 8 F^2 - 8 (x^2-t^2) F^3
\nonumber\\
&&\;\;
+ 4 (x^2-t^2)^2 F^4\biggr],
\eea
and $T_{00}$ is given by
\bea
\label{T00}
 T_{00} &=& 3(\partial_t W_0)^2 + (\partial_x W_1)^2 
- 2(\partial_t W_0)(\partial_x W_1) - (\partial_t W_1)^2
- (\partial_x W_0)^2
+ 2g[2W^2_1(\partial_t W_0)
\nonumber\\
&&\;\; - W_0 W_1(\partial_x W_0 + \partial_t W_1)]
+ 2g^2 W^2_1(W^2_1-W^2_0) + {\cal L}^{glue} 
\nonumber\\
& = & \frac{4}{g^2}\biggl[\frac{(x^2-t^2)(x^2-3t^2)}{x^2}
(\partial_x F)^2+ 4\frac{(x^2-3t^2)}{x}F \partial_x F+12F^2-16 x^2 F^3 
\nonumber\\
&&\;\;+8 x^2 (x^2-t^2) F^4\biggr] +{\cal L}^{glue}.
\eea
The field evolution depends on the initial 
conditions and we consider two possible scenarios, one based on the QCD 
instanton form and the other resembling the picture of Ref.~\cite{col}.

\subsection{Case I: Instanton-based model}

   In this subsection we use the instanton model for the initial conditions
to solve for the function $F(x^2)$. From Eqs.(\ref{inst},\ref{A_inst}),
the initial conditions for the two bubble walls at t=0 are
\bea
\label{initial}
  F(x,0)& = & \frac{1}{(x-3)^2 + \rho^2} + \frac{1}{(x+3)^2 + \rho^2}\\
\nonumber
  \partial_t F(x,0) & = & 0,
\eea
while the boundary conditions are
\bea
\label{boundary}
  F(-10,t) & = & F(10,t).
\eea
\begin{figure}
\centerline{\psfig{figure=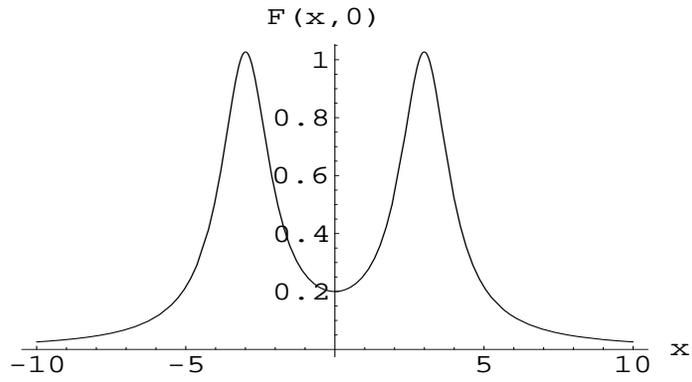,height=5cm,width=10cm}}
\caption{Initial condition, F(x,0), for the two bubbles with instanton
form}
\label{Fig.1}
\end{figure}
The initial conditions for $F(x,t)=F(x,0)$, given in Eq.(\ref{initial}) 
are shown
in Fig.~\ref{Fig.1}.

\begin{figure}
\centerline{\psfig{figure=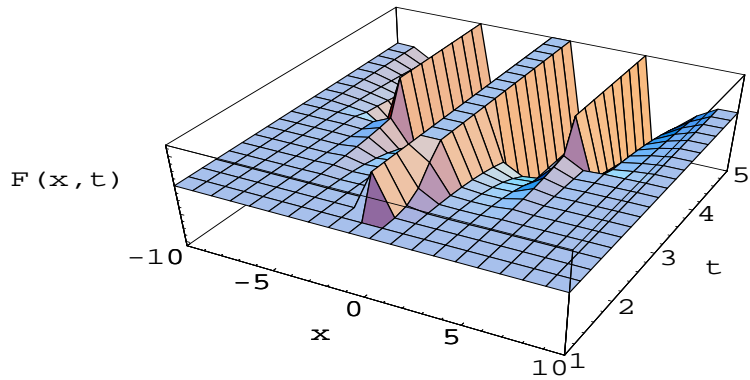,height=6cm,width=10cm}}
\caption{F(x,t) for the two bubbles with instanton-like form}
\label{Fig.2}
\end{figure}

\begin{figure}
\centerline{\psfig{figure=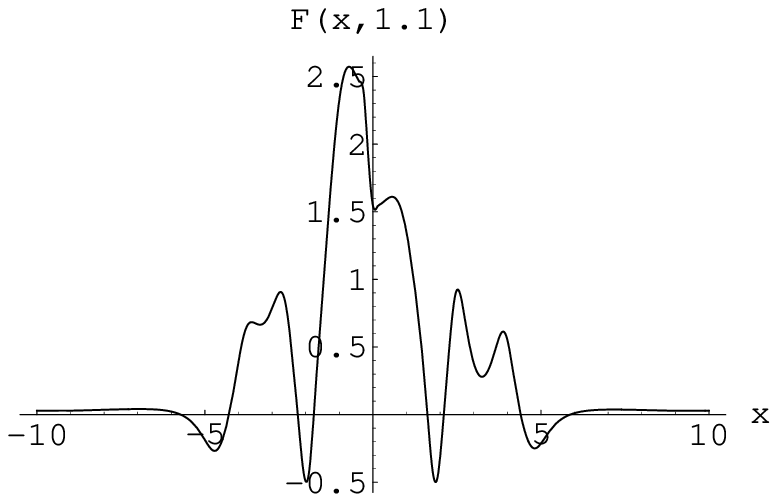,height=5cm,width=10cm}}
\caption{F(x,t) at t=1.1 for the two bubbles with instanton-like form}
\label{Fig.3}
\end{figure}

\begin{figure}
\centerline{\psfig{figure=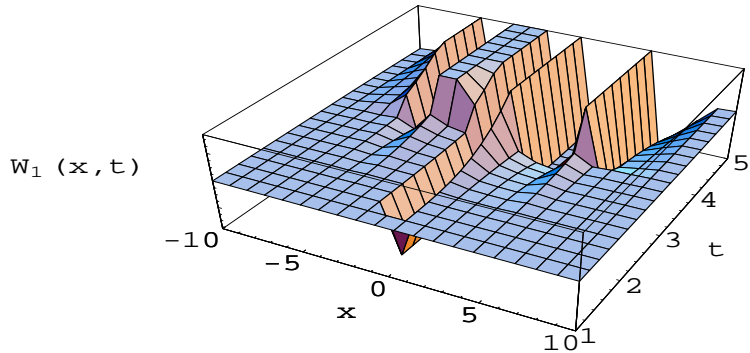,height=6cm,width=10cm}}
\caption{$W_1(x,t)=(2/g)xF(x,t)$ for the two bubbles with instanton-like form}
\label{Fig.4}
\end{figure}
The solution to Eq.(\ref{eominst}) is given in Fig. \ref{Fig.2}. Note the 
development of a gluonic wall at the x=0 collision region. The development
of the gluonic wall is clearly shown in Fig. \ref{Fig.3}, where F(x,t)
at time t=1.1 is shown. In comparison to F(x,0), shown in Fig. \ref{Fig.1}
one can see how the instanton-like bubbles have collided and an interior
gluonic wall produced. The wall is growing rapidly at this time, and
due to the singularities in the solution the accuracy of the calculations
for t $>$ 1.0 is limited, which is the origin of the violation of symmetry
about x=0 in Fig. 3 

The results for the function $W_1(x,t)=(2/g)xF(x,t)$, 
which is closely related to the color 
gluon field, are shown in Fig. \ref{Fig.4}. Note that the gluonic wall 
resembles a wall composed of instantons, as assumed in Ref. \cite{lsk2}. 

\begin{figure}
\centerline{\psfig{figure=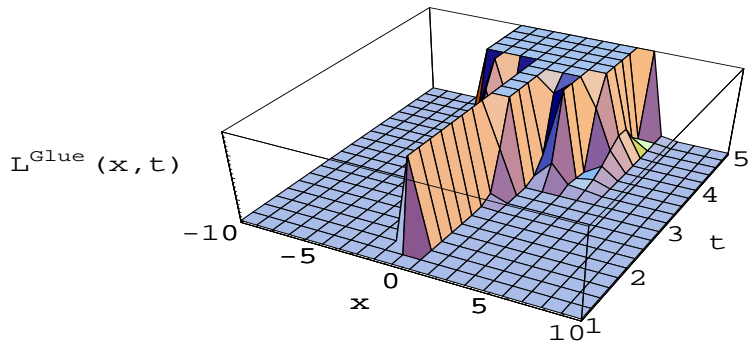,height=6cm,width=10cm}}
\caption{${\cal L}^{glue}$(x,t), the gluonic Lagrangian density for the 
two bubbles with  instanton-like form}
\label{Fig.5}
\end{figure}

\begin{figure}
\centerline{\psfig{figure=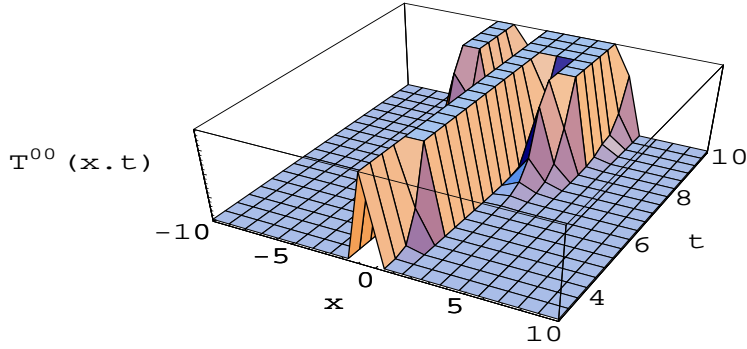,height=6cm,width=10cm}}
\caption{T$_{00}$(x,t), the energy density for the two bubbles with 
 instanton-like form}
\label{Fig.6}
\end{figure}

The gluonic Lagrangian density, ${\cal L}^{glue}$, for this solution
is shown in Fig. \ref{Fig.5}, and the energy density, $T_{00}$, is shown in
Fig. \ref{Fig.6}.

   In the light of the model calculations in Refs.~\cite{lsk2},\cite{lsk1}
these results are promising for studies of the CMBR correlations.

\subsection{Case II: Spontaneous symmetry breaking potential }
In 1+1 dimension, the instanton-like ansatz of the previous subsection
essentially takes the initial condition as $W_1(x,0)=x F(x,0)$ and
$W_0(x,0)=0$ and observes the evolution of the instanton wall.

In this subsection, we discuss another scenario, which contrast to the
instanton-like form but rather close to the
Coleman-like model of scalar $\phi^4$ field theory~\cite{col}. 
As we shall show below, the SU(2) color gauge field given by Eq.~(\ref{Lglue}) 
can be made to show the similar property of the scalar $\phi^4$ theory 
with the two degenerate vacua of the potential by taking 
$W_1(x,0)=w(x)$ and $W_0(x,0)=c$, where $c$ is a constant but not zero.
We should note that the above initial condition satisfies the gauge condition,
$\partial_x W_0(x,t)=\partial_t W_1(x,t)$ at $t=0$.

\vspace{5mm}

Then, the EOM given by Eq. (\ref{eom1}) becomes 
\bea\label{SSB_2}
\partial_x w(x)&=& g(w^2 - c^2),
\nonumber\\
\partial^2_x w(x) &=& 2 g^2 w(w^2 - c^2),
\eea
at $t=0$.  Note that the two coupled equations in Eq. (\ref{SSB_2})
give the same initial condition for the field $w(x)$, which
is obtained as 
\bea\label{SSB_3}
w(x)&=& -c\;{\rm tanh}[cg(x-x_0)],
\eea
where $x_0$ is an integration constant.
For this field configuration, the potential at $t=0$ can be easily
obtained as
\bea\label{SSB_4}
U(w)= \frac{g^2}{2}(w^2 - c^2)^2
\eea
up an additive constant.

We should note that the negative sign in Eq.~(\ref{SSB_3}) 
is uniquely determined from the 1st part of Eq.~(\ref{SSB_2}).
This is very different from the usual scalar $\phi^4$ theory 
model where only the 2nd part of Eq.~(\ref{SSB_2}) represents
the EOM of the theory and thus both signs could be solutions, i.e
$w(x)=\pm c\tanh[cg(x-x_0)]$. Our solution is not symmetric in x.

\begin{figure}
\centerline{\psfig{figure=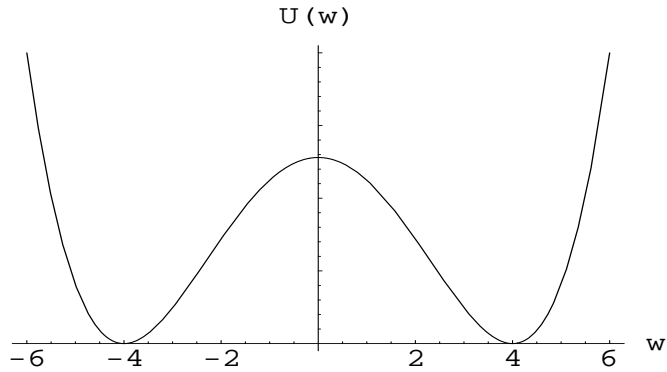,height=5cm,width=10cm}}
\caption{Potential energy $U(w)$ with the doubly degenerate 
minima, $w=\pm 4$ at $t=0$.\label{Fig7_fig}}
\end{figure}

Figure ~\ref{Fig7_fig} shows the potential as a function of $w$ for the
particular values of $c=4, g=1$ and $x_0=3$.  The potential has local minima
at $w=\pm 4$, which correspond to the degenerate vacuum states.  
  
\begin{figure}
\centerline{\psfig{figure=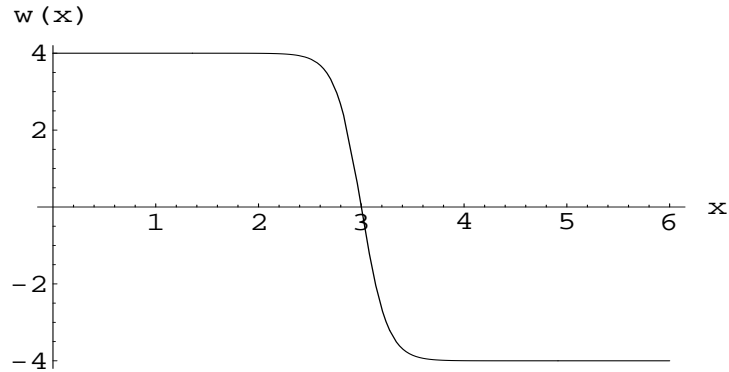,height=5cm,width=10cm}}
\caption{Initial($t=0$) bubble configuration with the bubble wall 
at $x=3$ where the collision occurs.\label{Fig8_fig}}
\end{figure}

Equation~(\ref{SSB_3}) is plotted in Fig.~\ref{Fig8_fig}.  This represents a
field configuration consisting of two regions of space (bubbles) separated
by a domain wall at $x=3$. We take this as the initial condition describing
a collision of a bubble in vacuum state $w=4$ on the left of the domain wall
with a bubble in vacuum state $w=-4$ on the right. Note that the two bubble
walls collide at x=3,t=0, and that in our 1+1 model the bubbles have
infinite radius. Thus our picture is a variation of the model of Ref\cite{col}.

Even though Eq.~(\ref{SSB_2}) does not allow solutions symmetric about 
$x=0$, rather than Eq.(\ref{SSB_3}) we use a symmetric ansatz for $w(x)$:
\bea
\label{wsym} 
w(x) &=& -c[1/2+{\rm tanh}[cg(x-x_0)]] -c[1/2-{\rm tanh}[cg(x+x_0)]] 
\eea
to
impose symmetric boundary condition, $W_1(-10,t)=W_1(10,t)$, as
well as $\partial W_i(x,t=0)/\partial t=0(i=0,1)$.
Otherwise, a non-symmetric boundary condition, such as $W_1(0,t)=-W(10,t)$
for the initial condition of Eq.~(\ref{SSB_3}),
does not give a solution to the EOM given by Eq.~(\ref{eom1}).
In the figures below we display only the solution for $x>0$, i.e. the 
solution with the correct initial $w(x)$ given by Eq.~(\ref{SSB_3}).

\begin{figure}
\centerline{\psfig{figure=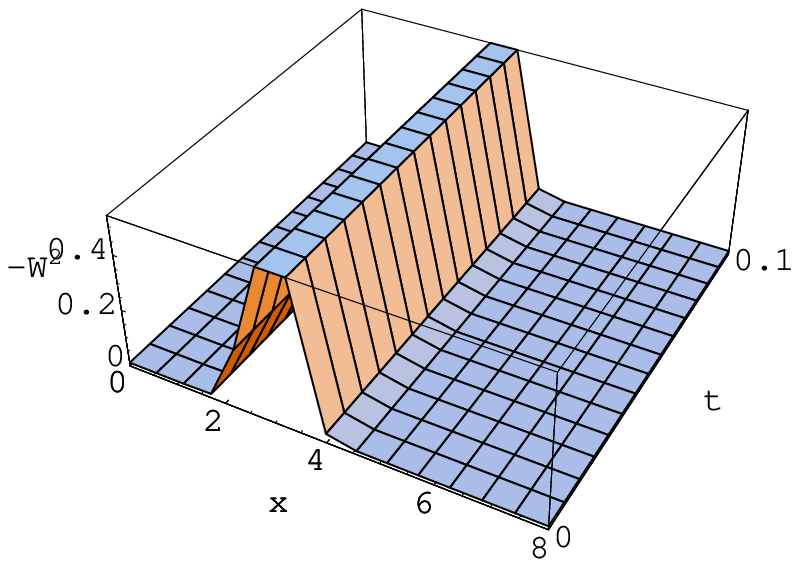,height=6cm,width=8cm}
\psfig{figure=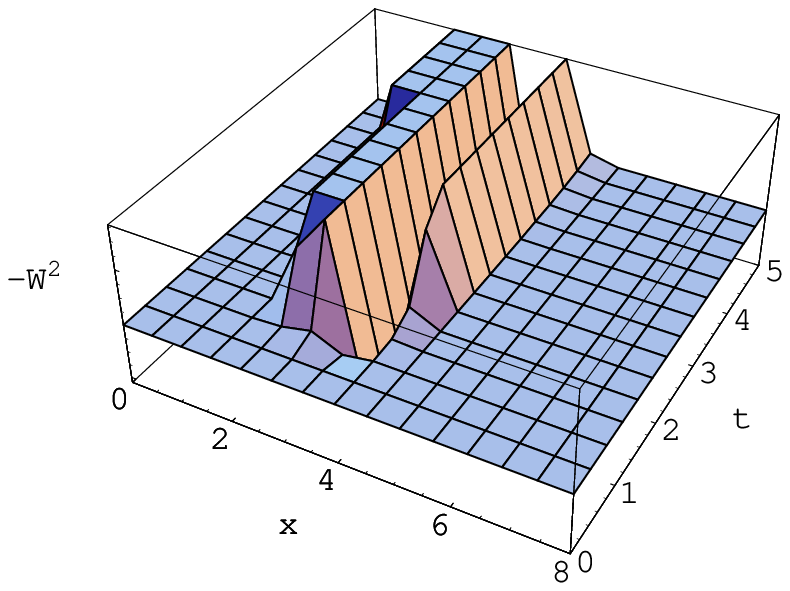, height=6cm,width=8cm}}
\caption{ Evolution of the gluon field squared,
$-W^2(x,t)=W^2_0(x,t)-W^2_1(x,t)$, with the domain 
wall-like form.\label{Fig9_fig}}
\end{figure}
The evolution of the gluon field squared, $-W^2(x,t)=W^2_0-W^2_1$, 
given by the solution to Eq.(~\ref{eom1}) with initial conditions
given by Eq.(~\ref{wsym}) in Minkowski
space, is shown in Fig.~\ref{Fig9_fig} for two different time intervals,
$0<t<0.1$(left) and $0<t<5$(right).
Although this domain-wall like ansatz is quite different from the
instanton ansatz, they show similar qualitative behavior of the 
field evolution.

\begin{figure}
\centerline{\psfig{figure=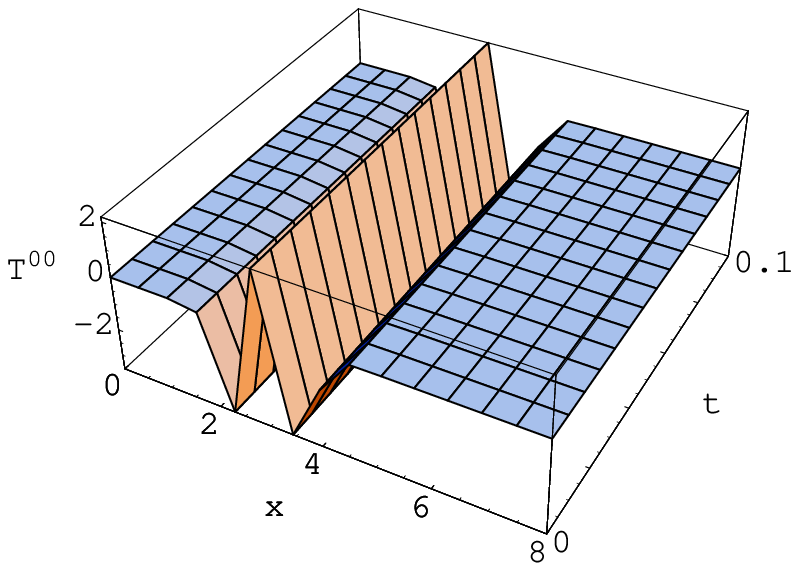,height=6cm,width=8cm}
\psfig{figure=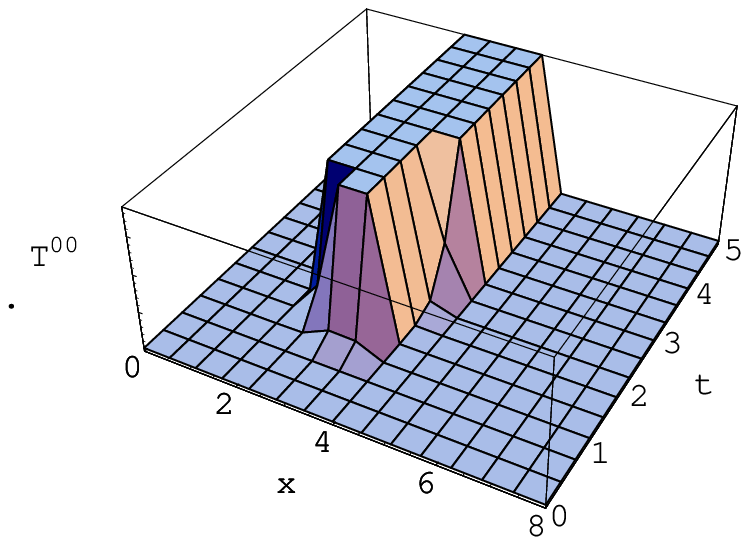, height=6cm,width=8cm}}
\caption{$T_{00}(x,t)$, the energy density with
the domain wall-like form.\label{Fig10_fig}}
\end{figure}

The energy density, $T_{00}(x,t)$ is shown in Fig.~\ref{Fig10_fig}
for two different time intervals, $0<t<0.1$(left) and $0<t<5$(right).
Again the energy density shown in Fig.~\ref{Fig10_fig} shows 
similar behavior to that in Fig.\ref{Fig.6} of instanton ansatz.
We also note that the gluonic Lagrangian 
density, ${\cal L}^{glue}$, is very similar to the energy density profile as
in the case of instanton-like form.


\section{Discussion and Conclusions}

   We have derived the equations of motion for a SU(2) color field theory,
motivated and guided by the instanton model of QCD. Solutions in one space
and one time dimension are studied as a simple picture for two QCD bubbles,
with an initial structure resembling that of an instanton wall for each
bubble, such as that assumed in Ref\cite{lsk2}. We indeed do find a gluonic 
structure evolving at the collision region, similar to that found in the
effective field calculations of Ref\cite{fz1}. In future research we shall
investigate the nucleation of the bubbles and the possible creation of
magnetic walls after the QCD phase transition for predictions of CMBR
correlations.
\vspace{.5 cm}

  This work was supported in part by the NSF grant PHY-00070888, in part 
by the DOE contract W-7405-ENG-36, and in part by the CosPA Project,
Taiwan Minestry of education 89-N-FA01-1-3. LSK and MBJ acknowledge 
partial support by the LANL Institute for Nuclear and Particle Astrophysics
and Cosmology, and HMC was supported in part by the Kyungpook National
University Research Fund, 2003.



\begin{thebibliography}{99}
\bibitem{lat1}S. Ijiri, Nucl. Phys B (Proc. Suppl.) {\bf 94}, 19 (2001).

\bibitem{lat2}Y. Iwasaki $et. al$, Z. Phys. C {\bf 71}, 343 (1996); 
C. DeTar, Nucl. Phys B (Proc. Suppl.) {\bf 42}, 73 (1995), 
T. Blum $et. al.$, Phys. Rev. D {\bf 51}, 5153 (1995).

\bibitem{nm}G. Neergaard and Jes Madsen, 
Phys. Rev. D {\bf 62}, 034005 (2000).

\bibitem{fz1} M.M. Forbes and A.R. Zhitnitsky, JHEP {\bf 0110}, 013 (2001);
 Phys. Rev. Lett. {\bf 85}, 5268 (2000); hep-ph/0102158 (2001).

\bibitem{lsk1} L.S. Kisslinger, hep-ph/0212206 (2002).

\bibitem{ign}J. Ignatius, K. Kajantie, H. Kurki-Suonio and M. Laine, 
Phys. Rev. D {\bf 50}, 3738 (1994); 
J. C. Miller and L. Rezzolla, Phys. Rev. D {\bf 51}, 4017 (1995); 
M.B. Christiansen abd J. Madsen, Phys. Rev. D {\bf 53}, 5446 (1996).

\bibitem{hms}S.W. Hawking, I.G. Moss and J.M. Stewart, 
Phys. Rev. D {\bf 26}, 2681 (1977).

\bibitem{col}S. Colman, Phys. Rev. D {\bf 15}, 2929 (1977); {\bf 16},1248(E);
C. Callan, S. Coleman, Phys. Rev. D {\bf 16}, 1762 (1977);
``Aspects of Symmetry'' (Cambridge University Press) (1985), Cpt. 7.

\bibitem{kv}T.W.B. Kibble and A. Vilenkin, Phys. Rev. D {\bf 52}, 679 (1995).

\bibitem{ae}J. Ahonen and K. Enqvist, Phys. Rev. D {\bf 57}, 664 (1998).

\bibitem{cst}E.J. Copeland, P.M. Saffin and 0.Tornkvist, 
Phys. Rev. D {\bf 61}, 105005 (2000).

\bibitem{nowak}E .Shuryak and I. Zahed, 
Phys. Rev. D {\bf 62}, 085014 (2000),
M.A. Nowak, E.V. Shuryak and I. Zahed ,
Phys. Rev. D {\bf 64}, 034008 (2001).

\bibitem{lsk2}L. S. Kisslinger, hep-ph/0202159 (2002).

\bibitem{bev} A.A. Belavin, A.M. Polyakov, A.S. Schwartz and Yu.S. Tyupkin,
Phys. Lett. B {\bf 59}, 85 (1975).

\bibitem{hooft}G. 't Hooft, Phys. Rev. D {\bf 14}, 3432 (1976).

\bibitem{ss} T. Sch\"{a}fer and E. V. Shuryak, 
Rev. Mod. Phys. {\bf 70}, 323 (1998).

\bibitem{chu}M-C. Chu and S. Schramm, Phys. Rev. D {\bf 51}, 4580 (1995);
 D {\bf 62}, 094508 (2000); 
T. Sch\"{a}fer and E.V. Shuryak, Phys. Rev. D {\bf 53}, 6522 (1996).  

\bibitem{Lin1} A. D. Linde, Nucl. Phys. B {\bf 421}, 421 (1983).

\end{thebibliography}
\end{document}